\documentclass[12pt]{article}
\pdfoutput=1
\usepackage{amsmath,amssymb,bbm}
\usepackage{epsfig}
\usepackage{slashed}
\usepackage{color}

\title{
\vspace{-2cm}
\vspace{3cm}
\bf \huge
Leptons in Composite MFV\vspace{1cm}}
\date{}
\author{{ Michele Redi\footnote{michele.redi@fi.infn.it}}\\
[10mm]
\normalsize\itshape  INFN, Sezione di Firenze, Via G. Sansone, 1; I-50019 Sesto Fiorentino, Italy
}

\begin{document}
\maketitle
\begin{abstract}
\medskip
\noindent
We study the lepton sector of composite Higgs models with partial compositeness. The standard anarchic scenario
is in conflict with the absence of observable charged lepton flavor violation. This tension can be completely solved in MFV 
scenarios that require either left-handed or right-handed SM leptons to be equally composite. Constraints on this scenario are weak and 
the composite lepton partners could be as light as few hundreds GeVs with interesting LHC signatures. The contribution to the muon $(g-2)$ in theories where the Higgs is a pseudo Nambu-Goldstone boson is also discussed.
\end{abstract}

\newpage
\tableofcontents

\section{Introduction}

The hope that new physics will be discovered at LHC relies on the existence of some special flavor
structure of the new degrees of freedom that does not jeopardize the special features of the Standard Model (SM), in particular suppression of flavor changing neutral currents and charged lepton flavor conservation.

The simplest hypothesis is that the flavor structure of the new physics is the same as in the SM itself, where flavor
symmetries are  only broken  by the Yukawa couplings. This possibility, known as Minimal Flavor Violation (MFV) \cite{bib:MFV},
is for example automatically realized in supersymmetric theories with gauge or anomaly mediation.
More recently it has been shown that MFV can also be  elegantly realized in theories where the Higgs is a composite state of new strong dynamics and  fermions are partially composite \cite{bib:weiler,bib:compositeMFV}. These models can also produce a different flavor structure, the one of ``anarchic models''  often considered in Randall-Sundrum scenarios, where a flavor protection against large new physics contributions is obtained because light generations are mostly elementary, suppressing their flavor transitions  \cite{bib:flavorconstraints}. 

The flavor protection in anarchic models is not perfect but overall the picture is plausible in the quark sector.
For leptons instead the absence of observable Charged Lepton Flavor Violation (CLFV) in present experiments poses a bigger challenge. In particular the bounds on the $\mu\to e \gamma$ transition strongly disfavor the anarchic hypothesis. With this motivation in mind, in this note we will  extend the construction of Ref. \cite{bib:compositeMFV}, to realize MFV in the lepton sector of partially composite Higgs models. 

Flavor problems are entirely solved for MFV leptons. Moreover the new fermions can be as light as few 
hundred GeVs, a possibility certainly inconsistent with the anarchic hypothesis. This is exciting because 
in the range below TeV, the heavy leptons could be produced at the LHC and some bounds can already be extracted
reinterpreting present searches. A correction to the $g-2$ of the muon is also generated that could potentially explain the 
discrepancy between the experimental value and the SM prediction. 

The paper is organized as follows. In section \ref{compositeleptons}, after reviewing the difficulties of  anarchic models, 
we present the MFV composite leptons scenario. In \ref{mfvsu2} we extend the MFV construction to theories with $SU(2)$ 
flavor symmetries \cite{bib:barbierisu2,bib:compositeMFV2}. Indirect and LHC bounds are considered in section \ref{bounds}. We discuss the contribution to the muon $(g-2)$ in section \ref{sec:gminus2}. We summarize in \ref{summary}. 

\section{Partially Composite Leptons}
\label{compositeleptons}

We work within the framework of composite Higgs models with partial compositeness. Fields can be divided
into elementary (SM) and composite. Each SM field mixes with composite states with 
identical quantum numbers under the SM gauge group that belong to a representation of the global symmetries of the composite sector.
We have in mind in particular the case where the Higgs is pseudo Nambu-Goldstone Boson (NGB) of a strongly coupled sector 
with symmetry $G$ spontaneously broken to $H$,  see \cite{bib:SILH} for the general framework. Many 
of the arguments concerning flavor however hold in general in scenarios with partial compositeness.

We will focus on the lepton sector. Assuming for simplicity the existence of a right-handed neutrino, the mixing terms are schematically
\begin{equation}
{\cal L}_{mixing}=\lambda_L \bar{l}_L L_R+ \lambda_{Re} \bar{\tilde{E}}_L e_R+ \lambda_{R\nu} \bar{\tilde{N}}_L\nu_R+h.c.
\end{equation}
where miniscule and capital letters refer to elementary and composite fermions and flavor indexes are implicit.
The mass of the charged leptons is given by,
\begin{equation}
\label{masses}
m_e\approx \frac v {\sqrt{2}} \epsilon_L\cdot Y_e \cdot \epsilon_{Re} 
\end{equation}
where $\epsilon\sim \lambda/m$ (with $m$ the mass of the composite partner), $v=$ 246 GeV and $Y_e$ is the relevant composite sector coupling.
Analogous formulae would hold for Dirac neutrinos. Within the logic of elementary-composite sectors it appears
natural to realize Majorana neutrinos by adding a large mass for the elementary right-handed neutrino. In this  case one finds,
\begin{equation}
m_\nu\approx v^2 \epsilon_L\cdot Y_e \cdot \epsilon_{R\nu}\cdot M^{-1}\cdot \epsilon_{R\nu}^T \cdot Y_\nu^T \cdot \epsilon_L^T
\end{equation}
where $M$ is the Majorana mass matrix of $\nu_R$.\footnote{For other realizations of neutrino masses, see \cite{bib:ramanneutrino,bib:rattazzi}. Neutrinos will not play a role in what follows.}

For several purposes the strong sector can be described truncating the theory to the lightest modes. For concreteness we consider the following simplified model \cite{bib:continosundrum},
\begin{equation}
\label{compositeL}
-{\cal L}_{comp}=m_L \bar{L}L+ m_{\tilde{L}} \bar{\tilde{L}}\tilde{L}+  (\bar{L}H) \cdot Y_e\cdot \tilde{E}+  (\bar{L}\tilde{H}) \cdot Y_\nu\cdot \tilde{N}+\dots
\end{equation}
where
\begin{equation}
L=\left(\begin{array}{c}
N\\
E
\end{array}\right)~~~~~~~~~~~~~~~~~~~~~~~~~
\tilde{L}=\left(\begin{array}{c}
\tilde{N}\\
\tilde{E}
\end{array}\right)\,.
\end{equation}
are a vectorial copy of SM fermions. The dots stand for other fermions required by the global symmetries of the theory (minimally custodial symmetry $SU(2)_L\times SU(2)_R$) and higher order terms suppressed by the compositeness scale. It should be kept in mind that the lagrangian (\ref{compositeL}) can just be used to provide estimates. Moreover the interactions above do not take into account the NGB 
nature of the Higgs that is crucial for certain observables, in particular the $g-2$ of the muon \cite{bib:dipoles}.

\subsection{Anarchic Leptons}

We start reviewing the severe bounds that one obtains in anarchic scenarios from leptons \cite{bib:blechman,bib:grossman,bib:agashe,bib:rattazzi}. The assumption  is that the composite sector has no hierarchies but is characterized by a typical mass scale $m_\rho$ and a coupling $g_\rho$ associated for example to the spin-1 resonances. In what follows we will treat fermionic parameters $m_\psi$ and $g_\psi \sim Y^{ij}$ as independent.

Let us first consider dipole operators. These are generated at 1-loop in the strong sector coupling with coefficient,
\begin{equation}
 \frac e {16 \pi^2} \frac v {m_\psi^2}\, \left[ U_L^\dagger \cdot \epsilon_L\cdot \, \left(a_1 Y_e \cdot Y^{\dagger}_e+a_2 Y_\nu \cdot Y^{\dagger}_\nu\right) \cdot Y_e \cdot \epsilon_{Re} \cdot U_R\right]_{ij}\,\bar{e}_L^i \sigma^{\mu\nu} e_R^j\, F_{\mu\nu}
 \label{dipole}
\end{equation}
where $U_{L,R}$ are the rotation matrices to the mass basis determined by eq. (\ref{masses}) and $a_{1,2}$ are model dependent order one numbers. In general both electric and magnetic dipole moments are generated of similar size. The estimate parametrically agrees with the explicit computation using the lagrangian  (\ref{compositeL}) but robustly  follows from the hypothesis of partial compositeness. 

In anarchic scenarios one obtains the estimate for $\mu \to e \gamma$ dipole operators,
\begin{equation}
 \left(\frac {g_\psi} {4\pi}\right)^2\frac e {m_\psi^2}\, \left(\frac {\epsilon_L^\mu}{\epsilon_L^e} m_e\bar{\mu}_L \sigma^{\mu\nu} e_R\, + \frac {\epsilon_L^e}{\epsilon_L^\mu} m_\mu\bar{e}_L \sigma^{\mu\nu} \mu_R\right)F_{\mu\nu}
\end{equation}
The most favorable choice of mixings is \cite{bib:grossman,bib:rattazzi},
\begin{equation}
 \epsilon_L^i\sim \epsilon_R^i \sim \sqrt{\frac {m_i}{g_\psi\, v}}
\label{optimalmixings}
\end{equation}
from which it follows,
\begin{equation}
{\rm Br}(\mu\to e \gamma)\sim 5\times \left(\frac {g_\psi}{3}\right)^4 \times \left(\frac {3\,{\rm TeV}}{m_\psi}\right)^4 \times 10^{-8}\,.
\end{equation}
Comparing with the recent experimental bound ${\rm Br}(\mu\to e \gamma)< 6 \times 10^{-13}$ @ 90 C.L. \cite{bib:MEG}, this result is 
orders of magnitude too large unless the coupling is extremely weak. This is however in tension with the logic of a strongly coupled sector.

\begin{table}
\begin{center}
\begin{tabular}{c||c|c|c}
& L$_1$\,& L$_2$\,  & {\rm EXP}\,\\  \hline
{\rm Br}($\mu\to e \gamma$) 		& $5\cdot 10^{-8}$ & $10^{-6}$ & $5\cdot 10^{-13}$\\ 
{\rm Br}($\tau\to e \gamma$) 	& $10^{-9}$	&  $10^{-7}$& $3\cdot 10^{-8}$	 \\ 
{\rm Br}($\tau \to \mu \gamma$)  & $10^{-7}$ & $10^{-6}$ & $4\cdot 10^{-8}$\\  \hline
{\rm Br}($\mu \to 3 e)$  & $10^{-13}$ &$10^{-12}$&$  10^{-12}$	 \\ 
{\rm Br}($\tau \to 3 e)$	& $10^{-13}$	&  $10^{-12}$&$ 3 \cdot 10^{-8}$	 \\ 
{\rm Br}($\tau\to 3 \mu)$  & $5\cdot 10^{-11}$ & $10^{-12}$&$ 2 \cdot 10^{-8}$ \\ \hline
{\rm Br}($\mu\to e)_{\rm Ti}$ & $5\cdot 10^{-13}$ & $10^{-11}$&$ 5 \cdot 10^{-13}$
\end{tabular}
\end{center}
\caption{\small CLFV branching fractions in anarchic scenarios. We choose $m_\rho=m_\psi=3$ TeV and scan the couplings $Y_e^{ij}\sim g_\psi$ around the central value 3. L$_1$ corresponds to the ``optimal choice'', while L$_2$ to $\epsilon_L={\rm Diag}[0.01,0.02,0.025]$. EXP is the experimental limit.}
\label{anarchicscan}
\end{table}

Let us consider $\mu \to eee$. The leading contribution is due to the flavor violating $Z-$couplings. Without protected
representations one finds,
\begin{equation}
\frac {\delta g_L^{ij}}{g_L}\sim \left(\frac { g_\psi^2\,v^2}{m_\psi^2}+  \frac {g_\rho^2\,v^2}{m_\rho^2}\right)\, \epsilon_L^i \epsilon_L^j
\label{modifiedcouplings}
\end{equation}
and similarly for the couplings to right-handed fermions. The two contributions arise perturbatively 
from the mixing of SM fermions and gauge fields with heavy partners, see for example \cite{bib:continosundrum}. 
From the first term, for the optimal choice of mixings (\ref{optimalmixings}) one finds,
\begin{equation}
{\rm Br}(\mu\to eee)\sim \left(\frac {g_\psi}{3}\right)^2\times  \left(\frac {3\, {\rm TeV}}{m_\psi}\right)^4\times10^{-13}
\end{equation}
while from the second,
\begin{equation}
{\rm Br}(\mu\to eee)\sim \left(\frac {g_\rho^2}{3\, g_\psi}\right)^2\times  \left(\frac {3\, {\rm TeV}}{m_\rho}\right)^4\times10^{-13}.
\end{equation}
This last contribution scales differently with the fermionic coupling so that reducing $g_\psi$ as demanded by $\mu \to e \gamma$ leads to problems 
with $\mu \to eee$ \cite{bib:blechman}.  A similar scaling is obtained for $\mu\to e$ transitions in nuclei, see \cite{bib:grossman}. 

These estimates can be validated performing a scan over the parameters of the model (\ref{compositeL}). In the table \ref{anarchicscan} we report the results for the optimal choice (L1) of mixings in eq. (\ref{optimalmixings}) and for left-handed mixings  of same order (L2), to reproduce hierarchies of the neutrino mixing matrix\footnote{Strictly this is not necessary as the large neutrino mixing angles  can originate from the neutrino sector.}. The results are in good agreement with the estimates.

\subsection{MFV Leptons}

The severe tension from CLFV provides a strong motivation to realize MFV in partially composite Higgs models.
We will follow the discussion for quarks in Refs. \cite{bib:compositeMFV,bib:compositeMFV2}. We  assume that the composite sector has an $SU(3)$ flavor symmetry with composite fermions transforming as triplets. This implies $Y_e= g_\psi\cdot{\rm Id}$ and degenerate masses for different flavors.
If the symmetry is respected by the left mixings,
\begin{equation}
\epsilon_L \propto {\rm Id} 
\end{equation}
the known mass hierarchies are entirely due to the right mixings, that will be proportional  to the SM Yukawa couplings
\begin{equation}
\epsilon_{Re}= \frac {y_e^{SM}}{\epsilon_L \, g_\psi}.
\end{equation}
The mixings, or equivalently the Yukawas, are the only sources of  breaking of  flavor symmetries as in the SM, therefore realizing the MFV hypothesis.
We call this scenario left-handed compositeness as the left-handed SM fermions have equal mixings. 

MFV can also be realized with universal right-handed mixing, $\epsilon_{Re}\propto {\rm Id} $.
The simplest option for neutrino masses in this case is to introduce a composite scalar triplet \cite{bib:rattazzi}. This allows
to write a bi-linear coupling of the elementary left-handed fields with the triplet that generates the dimension 5 operator
responsible for neutrino masses. The right-handed neutrinos and their partners are not necessary and 
MFV requires an $SU(3)$ global symmetry of the strong sector. Alternatively if neutrino masses are also generated through linear couplings 
MFV demands and $SU(3)^2$ flavor symmetry in right-handed compositeness \cite{bib:compositeMFV}.

\subsection{Beyond MFV}
\label{mfvsu2}

Realizing MFV in the lepton sector removes dangerous new physics flavor effects at least as long
as the see-saw scale $M \gg $ TeV. A variation of this can be obtained in theories 
based on $SU(2)$ flavor symmetries \cite{bib:barbierisu2,bib:compositeMFV2}, that allow to treat the
third generation independently, i.e. with different couplings and masses.  This is motivated in the 
quark sector by the heaviness of the top. These theories share  essentially the same success 
as MFV for what concerns flavor while they avoid  the strong bounds from precision tests and compositeness
of light quarks.

Once we assume $SU(2)$ to be a symmetry of the strong sector also composite leptons will be multiplets of this symmetry.
A natural choice is that the first two generations are doublets and the third singlets. For the strong sector couplings this implies
a structure,
\begin{equation}
Y_e= {\rm Diag}\left[g_\psi^1\,,g_\psi^1\,,g_\psi^2\,\right]
\end{equation}
and similarly for the masses. As before we can assume that the flavor symmetry is respected by the mixings. One can choose the basis,
\begin{eqnarray}
&&\epsilon_L={\rm Diag}\left[\epsilon_1\,,\epsilon_1\,,\epsilon_2\,\right]\nonumber \\
&&\epsilon_{Re}= U\cdot \hat{\epsilon}_{Re}
\end{eqnarray}
where $U$ is a unitary matrix and $\hat{\epsilon}_{Re}$ is diagonal.
The global symmetry of the theory in the charged sector is $SU(2)_L \times SU(3)_{e_R}$ broken by $\epsilon_{Re}$.

The reduced symmetry allows for new flavor structures and new physics flavor effects are not entirely decoupled.
In particular, due to the non-degeneracy of third generation masses and couplings CLFV is generated but one still obtains a suppression of flavor transitions, see \cite{bib:isidori} for the analog in supersymmetry. We consider two representative examples,
\begin{center}
L$^{SU(2)}_1$:~~~~~~
\begin{tabular}{|c|c|c|c|c|c|c|c}
\hline
$m_\rho$~(TeV) & $g_\rho$ & $\hat{\epsilon}_{L}$  \\ 
\hline 
$3$ & $3$ & $(0.05,0.05, 0.1)$  \\ \hline
\end{tabular}
\end{center}
\begin{center}
L$^{SU(2)}_2$:~~~~~~
\begin{tabular}{|c|c|c|c|c|c}
\hline
$m_\rho$~(TeV) & $g_\rho$ & $\hat{\epsilon}_{L}$ \\ 
\hline 
$3$ & $3$ & $(0.01,0.01, 0.1)$  \\ \hline
\end{tabular}
\end{center}

\begin{table}
\begin{center}
\begin{tabular}{c||c|c|c}
 & L$^{SU(2)}_1$\,& L$^{SU(2)}_2$\, & {\rm EXP} \\  \hline
{\rm Br}($\mu\to e \gamma$) 		& $10^{-9}$ & $ 10^{-12}$ &$5\cdot 10^{-13}$ \\ 
{\rm Br}($\tau\to e \gamma$) 	& $10^{-8}$	&  $10^{-9}$ & $3\cdot 10^{-8}$	 \\ 
{\rm Br}($\tau \to \mu \gamma$)  & $10^{-8}$ & $10^{-10}$ & $4\cdot 10^{-8}$ \\  \hline
{\rm Br}($\mu \to 3 e)$  & $10^{-12}$ &$10^{-15}$&$  10^{-12}$	 \\ 
{\rm Br}($\tau \to 3 e)$	& $10^{-10}$	&  $10^{-12}$&$ 3 \cdot 10^{-8}$	 \\ 
{\rm Br}($\tau\to 3 \mu)$  & $10^{-10}$ & $10^{-12}$& $2 \cdot 10^{-8}$ \\ \hline
{\rm Br}($\mu\to e)_{\rm Ti}$ & $10^{-11}$ & $10^{-14}$ & $ 5 \cdot 10^{-13}$
\end{tabular}
\end{center}
\caption{\small CLFV branching fractions in $SU(2)$ models for $m_\rho = m_\psi = 3$ TeV. The 
couplings $g_\psi^1$ and $g_\psi^2$ are scanned independently around the central value 3.}
\label{su2scan}
\end{table}

Estimates for CLFV obtained scanning over parameters of the model are reported in Table \ref{su2scan}.
The most delicate observable is again {\rm Br}($\mu\to e \gamma$) that is generated due to the difference
between $y_e^1/m_{\psi_1}$ and $y_e^2/m_{\psi_2}$ in eq. (\ref{dipole}). 
Flavor transitions are suppressed for hierarchical left mixings and can be in agreement with the data
for $\epsilon_1/\epsilon_2 < 0.1$ at least for fermions around 3 TeV.

\section{Bounds}
\label{bounds}

In this section we discuss the bounds and experimental signatures of composite leptons in MFV scenarios.
We focus on left-handed compositeness, similar arguments can be applied to right-handed compositeness.

\subsection{Compositeness and Precision Tests}

We start with the bounds on effective 4-Fermi operators. A list of experimental results can be found in \cite{pdgcompositeleptons}. 
With the normalization,
\begin{equation}
\frac {2\pi}{\Lambda^2} (\bar{f} f)^2
\end{equation}
the bound on $\Lambda$ can be up to 10 TeV. In our case these operators are generated 
by exchange of spin-1 resonances.  The strongest constraint arises from the effective operator $(\bar{l}_L \gamma^\mu l_L)^2$
whose coefficient  scales as,
\begin{equation}
\frac {g_\rho^2}{m_\rho^2} \epsilon_L^4
\end{equation}
From this we derive,
\begin{equation}
\epsilon_L^2<\frac 1 {4\, g_\rho} \times \frac {m_\rho}{{\rm TeV}} 
\label{4fermibound}
\end{equation}
Searches of ``excited leptons'' place a limit on the operator,
\begin{equation}
\frac 1 {2\,\Lambda} \bar{l'}_R \sigma^{\mu\nu}\left[g \frac {\tau^a} 2 W_{\mu\nu}^a+g' \frac Y 2 B_{\mu\nu} \right]\,l_L 
\label{excitedleptons}
\end{equation}
A conservative estimate in our scenario is \cite{bib:strongsignatures},
\begin{equation}
\frac 1 {\Lambda} \sim \epsilon_L  \left(\frac {g_\psi} {4\pi}\right)^2 \frac 1 {m_\psi} 
\end{equation}
so that $m_\psi \ll \Lambda$. Ref. \cite{bib:excitedleptons}  presents bounds in the plane 
$(m_\psi\,,\Lambda)$. For $m_\psi < \Lambda$ an approximate bound is $\Lambda > 11\,{\rm TeV}- 5\, m_\psi$ 
from which one derives,
\begin{equation}
\epsilon_L <\frac 1 {10} \left(\frac {4\pi}{g_\psi} \right)^2 \times \frac {m_\psi}{\rm TeV}\,.
\end{equation}

The bounds above are rather weak. Stronger constraints on left-handed compositeness can be derived
from precisions tests. In particular the coupling of the $Z$ to muons is measured with per mille precision,
\begin{equation}
R_h  =\frac {\Gamma(Z \to q \bar{q})} {\Gamma(Z\to \mu^+ \mu^-)}=20.0767\pm 0.25.
\end{equation}
Barring cancellations this implies\footnote{Tree level contributions could actually be avoided
using protected representations routinely used in the quark sector. In right-handed compositeness
this is automatic as long as $e_R^i$ couple to singlets \cite{bib:compositeMFV}.},
\begin{equation} 
\frac {\delta g^L_{Z\mu^+\mu^-}}{g^L_{Z\mu^+\mu^-}}< 0.002
\end{equation}
From the modified couplings (\ref{modifiedcouplings}) one finds,
\begin{equation}
\epsilon_L<  \frac 1 {5\,g_\psi}\times \frac {m_\psi}{{\rm TeV}}
\label{boundhadronicwidth}
\end{equation}
where we assume that the fermion contribution dominates.
This bound is similar to one on the mixing of the $b$ quark with the difference that the mixing 
can be small for leptons.  Reproducing the $\tau$ mass indeed requires,
\begin{equation}
\epsilon_L > \frac 1 {100\, g_\psi}
\end{equation}
where the lower bound corresponds to a fully composite $\tau_R$. 

Interestingly no bounds follow from lepton flavor universality also measured with per mille precision by LEP, 
because the shift of the couplings is universal. This is different in the theories based on $SU(2)$ flavor symmetries. 
The strongest bound follows from,
\begin{equation}
\frac {\Gamma(Z \to e^+ e^-)} {\Gamma(Z \to \tau^+ \tau^-)}=0.998\pm 0.003\,.
\end{equation}
from which we derive,
\begin{equation}
v^2 \Delta \left[\frac {g_\psi^2 \epsilon_L^2}{m_\psi^2}\right]_{e\tau}< 5\times 10^{-3}\,.
\end{equation}
Assuming a smaller compositeness for the first two generations this places a bound on the compositeness of $\tau_L$.

Effects on the $T-$parameter are also generated but they are small unless leptons are strongly composite. 
For example if the  the right-handed quarks couple to $SU(2)_R$ doublets the right-handed mixing breaks 
custodial symmetry and  one finds \cite{bib:SILH},
\begin{equation}
\hat{T}\sim \frac {g_\psi^4 \epsilon_R^4}{16 \pi^2} \frac {v^2}{m_\psi^2}
\end{equation}
which could be relevant if $\tau_R$ is strongly composite.

Let us briefly discuss Higgs physics. In anarchic models where the Higgs is a pseudo-NGB the corrections to SM rates are  
typically small, of order  $v^2/f^2$ where $f$ is the scale of the global symmetry breaking. In MFV one might expect 
larger  effects to $h\to \gamma\gamma$  and other observables with light leptons.
For example there is a correction to the coupling $h\gamma\gamma$ \cite{bib:azatov},
\begin{equation}
\frac{\delta g_{h\gamma\gamma}}{g_{h\gamma\gamma}}\sim 3\, \frac {g_\psi^2 v^2}{m_\psi^2}\,\epsilon_L^2 
\end{equation}
where we have included the multiplicity factor for 3 generations. Using the bound from precision tests
we get at most a per cent correction, irrelevant for LHC. Things could be different in the scenario
with right-handed compositeness where $\epsilon_R$ could be larger, see \cite{bib:perez}.

\subsection{LHC searches}
\label{lhc}

In the anarchic scenario lepton partners are expected to be definitely out
of reach for the LHC and indeed no specific searches have been performed so far.
In models with MFV the fermions can be light and can be searched at the LHC. 

We here consider the type of bounds that can be extracted from present searches.
The detailed phenomenology depends on the symmetries of theory and the representations of the new fermions and is beyond the scope
of this work. For the purpose of this section we will use the leading order lagrangian in eq. (\ref{compositeL}) with a vectorial generation of SM leptons. This is analogous to the renormalizable models considered in \cite{bib:strumia,bib:dermisek}.
\footnote{For small mixing the mapping between the models in Refs. \cite {bib:strumia,bib:dermisek} and eq. (\ref{compositeL}) is given by $Y_e \epsilon_L\approx\lambda_L$, $Y_e \epsilon_R\approx \lambda_E$, $Y_e\approx \lambda_{LE}\equiv \bar{\lambda}_{LE}^*$. In partial compositeness one parameter less is present because the SM Yukawa (\ref{masses}) is determined by the mixings and strong sector coupling.}

\begin{figure}[t!]
\begin{center}
$$\includegraphics[width=0.6\textwidth]{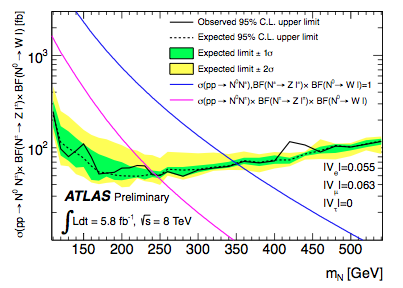}$$
\caption{\small ATLAS exclusion of type III see-saw models \cite{bib:ATLAStypeIII}.
}
\label{Fig:ATLAStypeIII}
\end{center}
\end{figure}

The most relevant LHC search is the one for type 3 see-saw models in Fig. \ref{Fig:ATLAStypeIII}  that can be reinterpreted
for the lepton partners. In type III see-saw one introduces a fermion triplet of $SU(2)_L$. In Weyl notations, neglecting hyper-charge, 
the electro-weak interactions are,
\begin{eqnarray}
&&g W_\mu^+\left[\bar{N}_0\bar{\sigma}^\mu N_- - \bar{N}_+\bar{\sigma}^\mu N_0\right]+ h.c.\nonumber \\
&&+ g W_\mu^3 \left[\bar{N}_+\bar{\sigma}^\mu N_+ - \bar{N}_-\bar{\sigma}^\mu N_- \right]
\label{typeIIIint}
\end{eqnarray}
and the decays widths satisfy \cite{bib:franceschini1},
\begin{eqnarray}
&\Gamma[N_0 \to W l]\approx 2  \Gamma[N_0 \to Z \nu]\approx 2 \Gamma[N_0 \to h \nu]\nonumber \\
&\Gamma[N_+\to W^+ \nu]\approx 2 \Gamma[N_+\to Z l^+]\approx 2 \Gamma[N_+\to h l^+]\,.
\label{pcint}
\end{eqnarray}
These formulae receive important corrections for light fermions but they will be sufficient for our estimates.

In our model, for the left-handed partners, the interactions are
\begin{eqnarray}
&& \frac g {\sqrt{2}} W_\mu^+\left[\bar{N}_L\bar{\sigma}^\mu E_L - \bar{E}_R^c\bar{\sigma}^\mu N_R^c\right]+ h.c. \nonumber \\
&&+ \frac g 2 W_\mu^3 \left[\bar{N}_L\bar{\sigma}^\mu N_L - \bar{E}_L\bar{\sigma}^\mu E_L- \bar{N}_R^c\bar{\sigma}^\mu N_R^c + \bar{E}_R^c\bar{\sigma}^\mu E_R^c \right]
\label{lsu2couplings}
\end{eqnarray}
The partners of right-handed electrons ($\tilde{E}$) only interact with hypercharge so we expect their production to 
be suppressed and we will not consider them. For the decay we have \cite{bib:continosundrum},
\begin{equation}
\Gamma[E\to  Z l]=\Gamma[E\to h l]=\frac 1 2 \Gamma[N\to W l]\approx\frac {g_\psi^2 \epsilon_R^2}{32\pi} m_\psi
\end{equation}
and we assume no other decay channels to be relevant (for example to $\tilde{E}$ or decay through dipole interactions in eq. (\ref{excitedleptons}), see Ref. \cite{bib:strongsignatures}). Note that in MFV the fermion partners are degenerate and will decay into the SM states of identical flavor.

The most sensitive search for our purposes is the one by ATLAS of type III 
see-saw models \cite{bib:ATLAStypeIII}. This search looks for $pp \to N_\pm N_0\to 4 l$ where 3 leptons come 
from the decay of $N_\pm\to Z l\to 3l$. This is sensitive to
\begin{equation}
\sigma(pp \to N_\pm N_0)\times {\rm Br}(N_\pm \to Z l)\times Br(N_0 \to W\,l)
\end{equation}
In the type III see-saw model one finds,
\begin{equation}
{\rm Br}^{III}(N_\pm \to Z l)\times {\rm Br}^{III}(N_0 \to W\,l) \approx \frac 1 4 \times \frac 1 2
\end{equation}
while with doublets,
\begin{equation}
{\rm Br}^{MFV}(E_{L,R} \to Z l)\times {\rm Br}^{MFV}(N \to W\,l ) \approx \frac 1 2 \times  1
\end{equation}
From eqs. (\ref{typeIIIint}) and (\ref{pcint}) and including a factor of 2 due to the degeneracy of electron and muon partners
we find that $\sigma \times {\rm Br}$ is roughly four times as in type III see-saw. From the ATLAS exclusion plot in Fig. \ref{Fig:ATLAStypeIII} a bound around 300 GeV will apply.

Let us mention that other searches could potentially improve the LHC sensitivity as suggested in \cite{bib:franceschini1,bib:franceschini2}, 
where final states with lepton+jets were considered in the context of type III see-saw models. Supersymmetric searches with lepton+jets and missing energy could also be reinterpreted for composite leptons if the missing energy cuts are not too strong.
The detailed collider phenomenology of light composite leptons will appear elsewhere.

\section{Muon $g-2$}
\label{sec:gminus2}

Both in anarchic and MFV scenarios sizable contributions to flavor diagonal observables are generated. Of particular interest is the $g-2$ of
the muon whose experimental value is presently 3.5 $\sigma$ away from the SM value. From eq. (\ref{dipole}) follows the estimate,
\begin{equation}
\Delta a_\mu \sim \left(\frac {g_\psi} {4\pi}\right)^2 \frac {m_\mu^2}{m_\psi^2}
\end{equation}
where as customary $a_\mu= (g-2)_\mu/2$.  A sharp prediction of our realization of MFV is the correlation between 
electron and muon contributions. To leading order in the mixings the following MFV operator is generated,
\begin{equation}
\bar{l}_L y_e^{ij} H^c \sigma^{\mu\nu} e_R^j\, e F_{\mu\nu}
\end{equation}
that implies,
\begin{equation}
\Delta a_e  = \frac {m_e^2}{m_\mu^2} \Delta a_\mu
\label{electrondipole}
\end{equation}
This result is a general consequence of MFV if the leading operators are allowed but also holds approximately in
anarchic models. This correlation is of potential interest as it could be tested in future experiments \cite{bib:paradisi2}.

At face value, to reproduce the muon anomaly, $\Delta a_\mu \approx 2.8 \cdot 10^{-9}$, one finds
\begin{equation}
m_\psi \sim g_\psi\times  150 \,{\rm GeV}  
\label{gminus2}
\end{equation}
This rough estimate suggests that composite fermions should be very light to account for  $(g-2)_\mu$, in agreement
with explicit results in Randall-Sundrum scenarios \cite{bib:beneke} and 4D renormalizable models \cite{bib:strumia,bib:dermisek}. 
Importantly in those models a large effect for $g-2$ is correlated with a modified  branching fraction of the Higgs into muons of order 10 times the SM value, 
that will be soon tested by the LHC. Within our MFV construction exactly the same correction
will appear in the $h \tau\bar{\tau}$ coupling. This is already grossly excluded by LHC data.

When the Higgs is a NGB the situation is more subtle. These models are characterized by the global 
symmetry breaking scale $f> v$ and strongly constrained by the symmetries. The modification of the Higgs couplings is of 
order $v^2/f^2$ times a numerical factor that depends on the fermion representations and is small for
phenomenologically plausible values of $f$ (a relatively safe choice is $f  > $ 800 GeV). Essentially for the same reason 
also the effect on $\Delta a_\mu$ is  small, even for light fermion partners. Indeed parametrically the masses scale as \cite{bib:manual},
\begin{equation}
m_\psi \sim g_\psi f 
\end{equation} 
that  is in conflict with (\ref{gminus2}) since it would imply $f$ around the electro-weak VEV.

Explicit computations in models with Higgs NGB can be found in \cite{bib:dipoles}. The typical size of the contribution is,
\begin{equation}
\Delta a_\mu \sim \frac 1 {16\pi^2} \times \frac {m_\mu^2}{f^2}
\end{equation}
even for fermions lighter than $f$. 

A  larger effect however can be obtained in certain regions of parameters. For example in the model described 
in \cite{bib:dipoles} an enhanced contribution is generated if the splitting between the masses of the composite fermions is small.
One finds
\begin{equation}
\Delta a_\mu \sim\frac 1 {16 \pi^2} \times \frac {m_\psi}{g_\psi f}  \times \frac {m_\mu^2}{f^2}
\label{amuNGB}
\end{equation}
where we have introduced the coupling $g_\psi = \Delta m/f$ that controls the SM Yukawa couplings.
The required effect could be obtained for $m_\psi\sim 10 \,g_\psi f$ for $f=800$ GeV.
More in general the dependence of these results on the composite fermion representations 
can be studied using the techniques of \cite{bib:dipoles}.

Let us briefly discuss electric dipole moments (EDM) . In anarchic models contributions to electric and magnetic dipole moments
are generated through loops of composite states and are expected of similar size. When the $(g-2)_\mu$ anomaly 
is explained, in light of (\ref{electrondipole}) the contribution to the electron EDM should be suppressed by a factor $10^{-3}$ compared
to the naive estimate to satisfy the experimental limit. As explained in \cite{bib:dipoles} in composite MFV scenarios
the contribution from loops of composite fermions can be aligned with the phase of leptons masses so that it contributes only 
to magnetic dipoles. Even so one should still worry about UV contributions from CP violating operators in the strong sector.
These however are not present under the minimal assumption that the strong sector respects CP. In this case
the physical phases are as in the SM and higher order effects could be sufficiently small.

\section{Summary}
\label{summary}

The hypothesis of partial compositeness allows new attractive realizations of flavor that are relevant for composite Higgs models. 
We have studied in this paper the leptonic sector.

\vspace{.5cm}

Summarizing the various possibilities\footnote{For other approaches see Refs. \cite{bib:agashe,bib:others}.}:

\begin{itemize}

\item{In anarchic scenarios the most important bound arises from the absence of observable $\mu \to e \gamma$ transitions. 
If the strong sector coupling is large as intuitively expected, then the branching fraction is several orders of magnitude above 
the experimental limit. Other regions of parameters lead to problems with other observables such as $\mu\to eee$. 
Overall these bounds appear to invalidate the anarchic hypothesis for leptons.}

\item{Realizing MFV completely solves the flavor problem in the lepton sector and contrary to quarks
does not have strong constraints from precision tests or compositeness. The lepton partners could be as light as few
hundreds GeVs and within the reach of the LHC with signatures similar to type III see-saw models. 
The new fermions contribute to the $(g-2)$ of the muon. We find that the contribution is too small in a large fraction of parameter space 
but there are  regions where it is conceivable to reproduce the long standing $(g-2)_\mu$ anomaly. 
When the latter is explained a contribution to $(g-2)_e$ is predicted that could be visible in future experiments. Problems with EDMs 
can be avoided if the strong sector respects CP.}

\item{One can extend the MFV paradigm to theories where the third generation is split from the others.
This is motivated in the quark sector by the heaviness of the top quark that suggests an approximate $SU(2)$ flavor symmetry.
Practically this interpolates between MFV and anarchic scenarios. In the lepton sector, beside $(g-2)_\mu$ and new fermions, 
charged lepton flavor violation could be visible in future experiments.}
\end{itemize}

\vspace{1cm}

\subsection*{Acknowledgments}
The work of MR is supported by the MIUR-FIRB grant RBFR12H1MW. I acknowledge the
Galileo Galilei Institute in Florence for the hospitality while this work was carried out. I am grateful to Roberto Franceschini 
for discussions on the phenomenology of heavy lepton partners and to Stefania De Curtis, Paride Paradisi and David Straub
for discussion of  related material.

\end{document}